\documentclass[twocolumn, pra, superscriptaddress]{revtex4-1}

\usepackage{graphicx}

\usepackage{amsmath}
\usepackage{amssymb}
\usepackage{amsthm}
\usepackage[usenames,dvipsnames]{color}
\usepackage[
 verbose,
 colorlinks=true,
 naturalnames=true,
 linkcolor=blue,
]{hyperref}
\usepackage{subfigure}

\usepackage{hyperref}
\hypersetup{colorlinks,linkcolor={blue},citecolor={blue},urlcolor={blue}}

	\def\be{\begin{equation}}
	  \def\ee{\end{equation}}
	  \def\ba{\begin{eqnarray}}
	  \def\ea{\end{eqnarray}}

\graphicspath{{"../Quantum Thermometry/Paper/"}}
	  
\begin{document}
\date{\today}
\title{Quantifying precision loss in local quantum thermometry via diagonal discord}

\author{Akira Sone}
\email{Both authors contributed equally to this work.}
\affiliation{Research Laboratory of Electronics, Massachusetts Institute of Technology, Cambridge, MA 02139, USA }
\affiliation{Department of Nuclear Science and Engineering, Massachusetts Institute of Technology, Cambridge, MA 02139}

\author{Quntao Zhuang}
\email{Both authors contributed equally to this work.}
\affiliation{Research Laboratory of Electronics, Massachusetts Institute of Technology, Cambridge, MA 02139, USA }
\affiliation{Department of Physics, Massachusetts Institute of Technology, Cambridge, MA 02139, USA}
\affiliation{4Department of Physics, University of California Berkeley, Berkeley, California 94720, USA}

\author{Paola Cappellaro}
\email{pcappell@mit.edu}
\affiliation{Research Laboratory of Electronics, Massachusetts Institute of Technology, Cambridge, MA 02139, USA }
\affiliation{Department of Nuclear Science and Engineering, Massachusetts Institute of Technology, Cambridge, MA 02139, USA}

\begin{abstract}
When quantum information is spread over a system through nonclassical correlation, 
it makes retrieving information by local measurements difficult---making global measurement necessary for optimal parameter estimation. In this paper, we consider temperature estimation of a system in a Gibbs state and quantify the separation between the estimation performance of the global optimal measurement scheme and a greedy local measurement scheme by diagonal quantum discord. In a greedy local scheme, instead of global measurements, one performs sequential local measurement on subsystems, which is potentially enhanced by feed-forward communication. We show that, for finite-dimensional systems, diagonal discord quantifies the difference in the quantum Fisher information quantifying the precision limits for temperature estimation of these two schemes, and we analytically obtain the relation in the high-temperature limit. We further verify this result by employing the examples of spins with Heisenberg's interaction.
\end{abstract}

\maketitle

\section{Introduction}
Quantum metrology~\cite{giovannetti2006,giovannetti2011advances,Degen16x} utilizes quantum resources such as entanglement and coherence to improve the precision of measurements beyond classical limits. The ultimate precision of estimating a parameter $\lambda$ from a quantum state $\rho(\lambda)$ is given by the quantum Cramer-Rao bound~\cite{Helstrom_1976,Holevo_1982,Yuen_1973}, which bounds the estimation variance $\delta \lambda^2\ge 1/\mathcal{F}\left(\lambda, \rho(\lambda)\right)$, by the quantum Fisher information (QFI):~
$
\mathcal{F}\left(\lambda, \rho(\lambda)\right)\equiv -2\lim_{\epsilon\to 0}{\partial^2\epsilon}\mathbb{F}\left[\rho\left(\lambda\right), \rho\left(\lambda+\epsilon\right)\right]
$, where $\mathbb{F}\left[\rho,\sigma\right]$ is the fidelity between states $\rho,\sigma$.

Applications range from clock synchronization~\cite{Giovannetti_2001}, to quantum illumination~\cite{Lloyd2008,Tan2008,Zhuang_2017}, superdense measurement of quadratures~\cite{Genoni_2013,steinlechner2013quantum,ast2016reduction} and range velocity~\cite{zhuang2017entanglement}, distributed sensing~\cite{ge2017distributed,proctor2018multiparameter,zhuang2018distributed}, point separation sensing~\cite{nair2016,lupo2016,Tsang_2016,kerviche2017}, and magnetic field sensing~\cite{Baumgratz_2016,Taylor08}. 

The most common sensing protocols aim at estimating parameters, with extension to quantum system identification, including Hamiltonian identification ~\cite{Sone17a, Wang16, Zhang15} and dimension estimation~\cite{Sone17b, Owari15}. 
All the schemes above can be seen as various kinds of channel parameter estimation, where the channels are given as a black box with unknown parameters. 
There are, however, other important sensing tasks that go beyond the framework of channel parameter estimation, most notably temperature estimation.

Temperature is an essential quantity in thermodynamics. As the study of thermodynamics extends to the nanoscale, temperature estimation also requires a fully quantum treatment~\cite{brunelli2011qubit, brunelli2012qubit,mehboudi2015thermometry,correa2015individual,raitz2015experimental,jevtic2015single,mancino2017quantum,Xie17,correa2017,Pasqualle17sequential, Paris16Landau, Hofer17}. Correa~\textit{et al.}~\cite{correa2015individual} showed that QFI for temperature estimation is proportional to the heat capacity $C(T)$. 
Then, the optimal measurement strategy involves projective measurements of the energy eigenstates, since heat capacity corresponds to energy fluctuations. Unfortunately, performing projective measurements of (global) energy eigenstates is typically hard, as eigenstates usually contain nonclassical correlations among different parts of the system. 

\begin{figure}
\includegraphics[width=5.5cm]{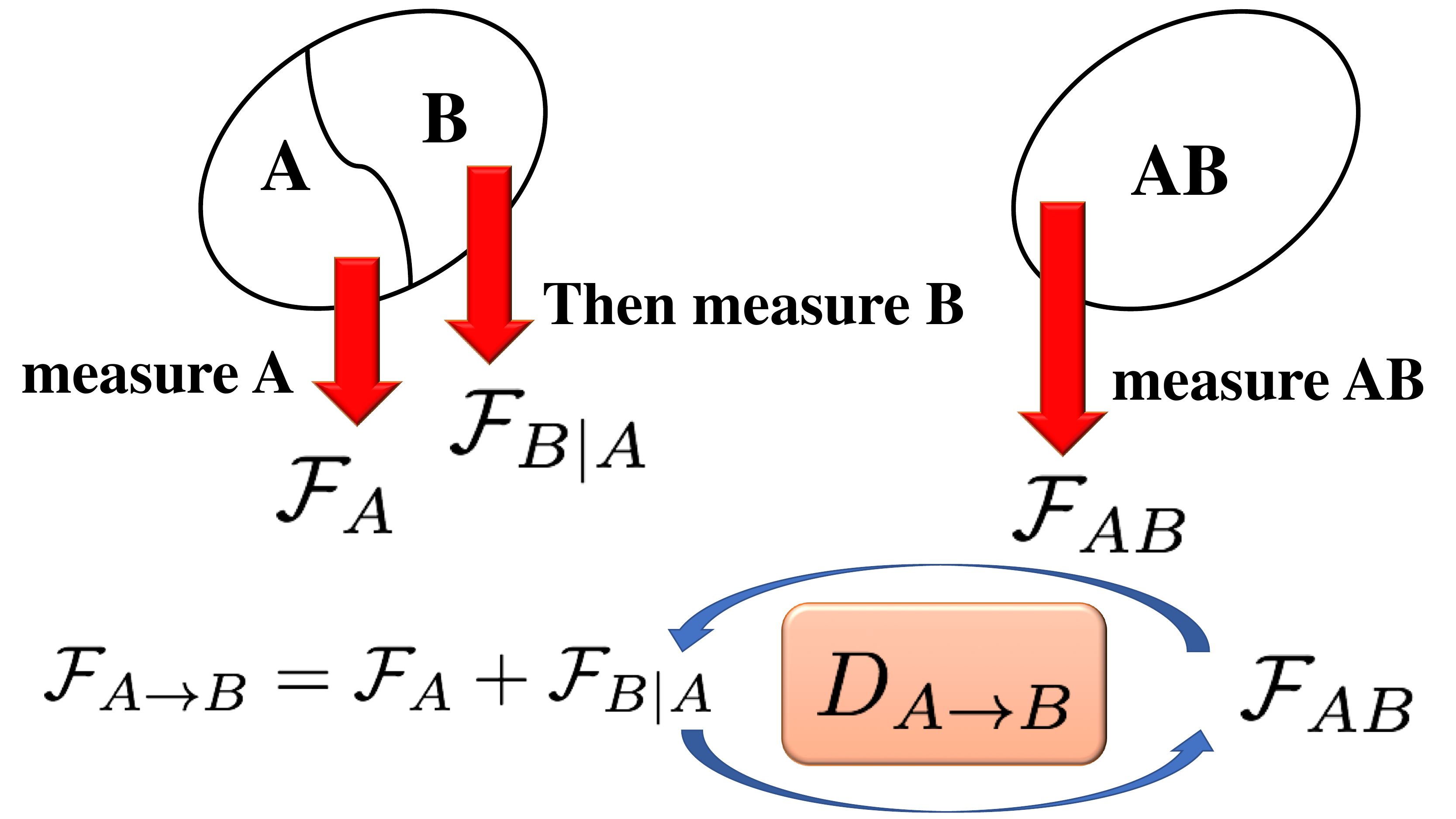}
\caption{Greedy local scheme: We first measure a subsystem $A$ and then measure the other subsystem $B$. The measurement on $A$ is optimum in the sense of local QFI. The constrained QFI of this greedy local scheme is given as $\mathcal{F}_{A\to B}(T)=\mathcal{F}_{A}(T)+\mathcal{F}_{B|A}(T)$, and we explore how the quantum discord $D_{A\to B}$ affects the loss of precision loss, i.e., $\mathcal{F}_{AB}(T)-\mathcal{F}_{A\to B}(T)$. 
}
\label{fig:idea}
\end{figure}

Recent works~\cite{Pasquale16, Palma17} considered measurements on a single subsystem, finding that the local QFI~\footnote{In Ref.~\cite{Pasquale16, Palma17}, they define the local quantum thermal susceptibility as the local QFI for estimating the inverse temperature.} bounds the ultimate achievable precision. We can however expect that a more general measurement scheme, with sequential local measurements on multiple subsystems and (classical) feed forward from previous measurements, could improve the estimate precision.  This scheme still remains practical and belongs to the class of local operations and classical communication (LOCC)~\cite{nielsen1999conditions}. {A practical LOCC protocol is the \textit{greedy} local scheme, where we sequentially measure each subsystem with a local optimal measurement (see Fig.~\ref{fig:idea}). We call the constrained QFI of the greedy local scheme the LOCC QFI.}

For systems with classical Gibbs states, given by product states among subsystems, such local greedy schemes are optimal. However, for generic quantum systems, Gibbs states can be highly nonclassical. Thus, temperature as a global property requires global measurements to be optimally estimated, while local sequential schemes cannot achieve optimal precision due to the nonclassical correlations in the system. The local QFI has been recently shown to depend on the correlation length at low temperature~\cite{Hovhannisyan17}. In a related metrology task, channel parameter estimation, the correlation metric for pure quantum states based on the local QFI, was shown~\cite{Kim17} to coincide with the geometric discord~\cite{dakic2010necessary}. {Also, the relation between the decreasing QFI due to the measurements on the total system and the disturbance has been considered~\cite{Seveso18}.}

In order to explore the relation between precision loss---the difference between QFI for the global measurement scheme and the LOCC QFI---and nonclassical correlation more broadly, we focus on temperature estimation and seek a relation between precision loss and quantum discord~\cite{Olivier01}, which quantifies nonclassical correlations in a quantum system.

{We focus on the high-temperature limit and analytically find that the precision loss can be exactly quantified by a quantum correlation metric in this regime, despite that entanglement or nonclassical correlations are expected to play lesser roles. }
In addition, temperature estimation at high temperature is a practically important task as the capability of performing coherent operations at room temperature is a desirable feature for quantum information processing devices. Also, quantum phenomena such as superconductivity~\cite{keimer2015quantum,fradkin2015colloquium} survive at temperatures as high as 165 K.

In this paper, we explore the contribution of nonclassical correlations to the ultimate precision limit of temperature estimation by comparing a greedy local scheme (see Fig.~\ref{fig:idea}) to the optimal global measurement on the total system. We prove that for a bipartite system in the Gibbs state at high temperature, {precision loss defined in terms of} QFI is quantified by the diagonal discord~\cite{Liu17}, which is the upper bound of the quantum discord and recently has been shown to play an important role in thermodynamic processes such as energy transport~\cite{Lloyd15discord}. We further generalize this relation to multipartite systems, showing that the precision loss is quantified by a multipartite generalization of the diagonal discord. 

\section{Thermometry in Bipartite Systems}
Consider temperature estimation from a Gibbs state $\rho=\mathcal{Z}^{-1}\exp(- H/T)$ at temperature $T$, where $H$ is the Hamiltonian of the system, $\mathcal{Z}\equiv \text{Tr}[\exp(- H/T)]$ is the partition function, and we set the Boltzmann constant $k_B=1$. QFI is given as~\cite{correa2015individual} (see also Appendix~\ref{app:QFIGibbs})
\begin{equation}
\mathcal{F}(T)={C(T)}/{T^2}.
\label{eq:Fisher1}
\end{equation}
Given $C(T)=\partial_T\langle H\rangle=\delta H^2/T^2$, energy measurement---projection to energy eigenstates---is optimal. However, global measurements are usually hard to implement. {The more practical way is to estimate the temperature $T$ by measuring a subsystem. Suppose that a bipartite system is composed of subsystems $A$ and $B$, and we measure $A$.} The local QFI $\mathcal{F}_{A}(T)\equiv\mathcal{F}\left(T,\rho_A\right)$, where $\rho_A=\text{Tr}_{B}(\rho_{AB})$, quantifies the ultimate precision limit of any possible \textit{local} measurement on a single subsystem $A$. Since the reduced state $\rho_A$ is usually not a Gibbs state,  $\mathcal{F}_{A}(T)$ does not follow Eq.~(\ref{eq:Fisher1}).

In addition to measurement on a subsystem $A$, one can proceed to perform measurement on the reminder of the system, $B$, in order to estimate the temperature. In the greedy local scheme (see Fig.~\ref{fig:idea}), the measurement on $A$ is the local optimum measurement operators $\{M_x\}_A$. Then, the quantum state of $B$ conditioned on measurement result $x$ is
$
\rho_{B|M_x}={1}/{p_x}\text{Tr}_A \left[
\left(M_x\otimes \openone_B\right) \rho_{AB}
\left(M_x^\dagger\otimes \openone_B \right) 
\right],
$
with $p_x=\text{Tr} \left[(M_x\otimes\openone_B)\rho_A (M_x^\dagger\otimes\openone_B)\right]$ the measurement probability. The conditional local QFI is given by $\mathcal{F}_{B|M_x}(T)=\mathcal{F}(T,\rho_{B|M_x})$ and the {unconditional QFI} is $\mathcal{F}_{B|A}(T)=\sum_x p_x(T) \mathcal{F}_{B|M_x}(T)$. Note that the measurement achieving $\mathcal{F}_{B|M_x}(T)$ may depend on $x$; thus, feed forward is required. 
The LOCC QFI $\mathcal{F}_{A\to B}$ from the above consecutive local optimal measurements on $A$ and $B$ quantifies the precision of the local greedy temperature measurement protocol. {Then, the LOCC QFI can be written as
\begin{equation}
\mathcal{F}_{A\to B}(T)=\mathcal{F}_A(T)+\mathcal{F}_{B|A}(T),
\label{F_Markovian}
\end{equation}
which is derived from the additivity of Fisher information~\cite{Lu12, Micadei15}. (In Appendix~\ref{app:Markovian}, we also provide our proofs.)} 

By definition, $\mathcal{F}_{A\to B}(T)\le\mathcal{F}_{AB}(T)$, with equality satisfied for $\rho_{AB}$ in a product state. 
Then, the precision loss $\Delta \mathcal{F}(T)\equiv \mathcal{F}_{AB}(T)-\mathcal{F}_{A\to B}(T)$ is generally related to bipartite nonclassical correlations with a proper measure. Here in particular, we demonstrate a link to 
quantum discord. 

Let $I_{AB}$ be the quantum mutual information between $A$ and $B$: $I_{AB}=S_A+S_B-S_{AB}$, where $S_i=-\text{Tr}[\rho_i\ln \rho_i]$ is the entropy of the state $\rho_i$. Suppose that we measure subsystem $A$ with projective measurements $\{\Pi_{j}^{A}\}$ (i.e., $\Pi_{i}^A\Pi_{j}^A=\delta_{ij}\Pi_{j}^A$). The classical correlation is defined as 
$
J_{B|A}=S_B-\min_{\{\Pi_{j}^{A}\}} S_{B|\{\Pi_{j}^A\}},
$
with $S_{B|\{\Pi_{j}^A\}}=\sum_{j}p_j S_{B|\Pi_{j}^A}$, where $S_{B|\Pi_{j}^A}$ is the entropy of the postmeasurement state $\rho_{B|\Pi_{j}^A}$.
Then, the quantum discord of $\rho_{AB}$ as $A$ being measured is defined as $D_{A\to B}=I_{AB}-J_{B|A}$, or explicitly
\begin{equation}
D_{A\to B}=-S_{AB}+S_A+\min_{\{\Pi_{j}^A\}}S_{B|\{\Pi_{j}^A\}}.
\label{eq:discord}
\end{equation}
Suppose that, instead of performing the minimization, we choose $\Pi_j^A\equiv |j\rangle_A\langle j|$ as the eigenbasis of $\rho_A$ in Eq.~(\ref{eq:discord}), i.e., $\rho_A=\sum_{j}r_j|j\rangle\langle j|=\sum_j r_j\Pi_{j}^{A}$. In this case, it becomes the \textit{diagonal discord} $\mathcal{D}_{A\to B}$~\cite{Liu17}. Note that diagonal discord has an alternative expression $\mathcal{D}_{A\to B}=\inf_{\pi_A}S(\pi_A(\rho_{AB}))-S(\rho_{AB})$, where $\pi_A\equiv \sum_j\Pi_{j}^{A} \otimes \openone_B$ and $\inf$ is due to possible degeneracy of the eigenbases.

In the high-temperature limit, for the finite-dimensional bipartite systems in the Gibbs state at temperature $T$, we find that the precision loss is given by
\begin{equation}
\Delta\mathcal{F}(T)=-(1/T)\partial_T\mathcal{D}_{A\to B}(T)+O(T^{-5}).
\label{eq:discordFisher}
\end{equation}
This relation can be proved by realizing that in the high-temperature limit, the partial states are still well approximated by the Gibbs states. 
Then Eq.~(\ref{eq:Fisher1}) is still approximately valid and one can relate the local QFI to the entropy of the subsystem and thus to {diagonal discord}. Let us write the total Hamiltonian as 
\begin{align*}
H=H_A+H_B+H_{AB},
\end{align*}
where $H_A$ and $H_B$ are the system Hamiltonians of $A$ and $B$, respectively, and $H_{AB}$ is the interaction Hamiltonian between $A$ and $B$. The Gibbs state of the total system is then $\rho_{AB}=\mathcal{Z}_{AB}^{-1}\exp\big(-\beta (H_A+H_B+H_{AB})\big)$, where $\beta=1/T$.
From Eq.~(\ref{eq:Fisher1}), since  the heat capacity is given by $C_{AB}(T)=T\partial_T S_{AB}(T)$, we can write
\begin{equation}
\mathcal{F}_{AB}(T)=C_{AB}(T)/T^2=(1/T)\partial_T S_{AB}(T).
\label{eq:QFIAB}
\end{equation}

{For a general finite-dimensional system,} in the high-temperature limit $\beta\ll 1$, $\rho_{AB}$ can be written as 
{
\begin{equation}
\rho_{AB}=\frac{1}{d_{AB}}\Big(\openone_{AB}-\beta \Big(H-\frac{\text{Tr}[H]}{d_{AB}}\Big)\Big)+O(\beta^2).
\label{eq:highTexpansion}
\end{equation} 
}
Within the same approximation, the reduced state $\rho_A=\text{Tr}_B[\rho_{AB}]$ is $\rho_A\propto(\openone_A-\beta H_A-\beta\Omega_A)+O(\beta^2)$,
where {$\Omega_A=\text{const}+\frac{1}{d_B}\sum_{k}\langle E_k^{(B)}|H_{AB}|E_k^{(B)}\rangle$}, which is \textit{independent} of the temperature $T$ (here $E_k, |E_k^{(B)}\rangle$ are $B$'s energy eigenvalues and eigenstates). {Note that when the interaction between $A$ and $B$ is absent, i.e. $H_{AB}=0$, due to $[H_A,~H_B]=0$, the Gibbs' state of the total system can be written as the product Gibbs' state of the subsystems, which are only relevant to their system Hamiltonians. Therefore, in this case, we have $\Omega_A=0$ for any temperature $T$. In the high-temperature limit, in general,} $H_{A}^{\text{eff}}=H_A+\Omega_A$ behaves as an effective Hamiltonian for subsystem $A$. Therefore, at high temperature, $\rho_A$ is approximated by a Gibbs state, 
$\rho_A\simeq\mathcal{Z}_A^{-1}\exp\big(-\beta H_{A}^{\text{eff}}\big)$,
with $\mathcal{Z}_A\equiv \text{Tr}[\exp\big(-\beta H_{A}^{\text{eff}}\big)]$. Then, the local QFI still follows Eq.~(\ref{eq:Fisher1}) and can be written, within this approximation, as 
\begin{equation}
\mathcal{F}_A(T)\simeq(1/T)\partial_T S_A(T).
\label{eq:QFIA}
\end{equation}
The measurements that saturate this local QFI are the projectors $\Pi_j^A$ onto local eigenstates of $\rho_A$, since they are also eigenstates of the effective Hamiltonian $H_A+\Omega_A$. 

Similar to $\rho_A$, the conditional state $\rho_{B|\Pi_j^A}$ after measuring $A$ can be also approximated by a Gibbs state, $\rho_{B|\Pi_{j}^A}\simeq \mathcal{Z}_{B|\Pi_{j}^A}^{-1} \exp\big(-\beta H_{B|\Pi_{j}^A}^{\text{eff}}\big)$, with effective Hamiltonian $H_{B|\Pi_{j}^A}^{\text{eff}}=H_B+\Omega_{B|\Pi_{j}^A}$, where {$\Omega_{B|\Pi_{j}^A}=\text{const}+\langle j|H_{AB}|j\rangle$.} This allows us to relate the corresponding local QFI to entropy
\begin{equation} 
\mathcal{F}_{B|\Pi_j^A}(T)\simeq(1/T)\partial_T S_{B|\Pi_j^A}(T),
\label{eq:QFIB}
\end{equation}
where $S_{B|\Pi_j^A}(T)$ is the entropy of subsystem $B$ after the measurement $\Pi_j^A$. 
By selecting a set of projection measurements that minimize $B$'s entropy, we can relate the entropies to {diagonal discord}. More precisely,
let $\{\Pi_{j*}^A\}$ be the set of projection measurements on subsystem $A$ such that $\sum_{i}p_{i*}(T)S_{B|\Pi_{j*}^A}(T)=\min_{\{\Pi_{j}^A\}}\sum_{j}p_{j}(T)S_{B|\Pi_{j}^A}(T)$. 

From Eqs.~(\ref{eq:QFIAB}-\ref{eq:QFIB}), we have 
\begin{equation}
-\partial_T \mathcal D_{A\to B}(T)\simeq T\Delta\mathcal{F}(T)-\sum_{k}\partial_T p_{k*}(T) S_{B|\Pi_{k*}^A}(T).
\label{eq:diagonaldiscord}
\end{equation}
Note that for finite dimensional system we have (see Appendix~\ref{app:prob})
\begin{equation}
(1/T)\sum_{k}\partial_T p_{k*}(T)S_{B|\Pi_{k*}^A}(T)=O(T^{-5}).
\label{eq:prob}
\end{equation}
Then we have two cases. A trivial case is when the greedy local method is asymptotically optimal at high temperature, i.e., $\lim_{T\to\infty}\ \Delta \mathcal{F}(T)/\mathcal{F}(T)=0$, as the deviation $\Delta \mathcal F$ is no longer important.
If instead $\Delta \mathcal{F}(T)/\mathcal{F}(T)$ remains finite at high temperature, since QFI (see Appendix~\ref{app:fisherorder}) 
\begin{equation}
\mathcal{F}(T)\simeq O(T^{-4}) 
\label{eq:fisherorder},
\end{equation}
we must also have $\Delta \mathcal{F}(T)=O(T^{-4})$, which is then the dominant term in the right-hand-side of Eq.~(\ref{eq:diagonaldiscord}) and we recover 
Eq.~(\ref{eq:discordFisher}).

We can make these ideas more concrete by presenting an example given by a two-qubit X state~\cite{Ali10, Chen10, Yurischev11}. We consider the general Heisenberg interaction Hamiltonian $H=\left(1/2\right)\left({B_1}Z_A+{B_2}Z_B+{J_x}X_AX_B+{J_y}Y_AY_B+{J_z}Z_AZ_B\right)$, where $X_k$, $Y_k$, and $Z_k$ are the Pauli matrices acting on $k$th qubit. The Gibbs state of this system is the two-qubit $X$ state.
In the high-temperature limit, the quantum mutual information term is $-(1/T)\partial_T I_{AB}(T)=-\left({J_x^2+J_y^2+J_z^2}\right)/(4T^4)+O(T^{-5})$ and the classical correlation term $-(1/T)\partial_T J_{B|A}(T)={J_z^2}/(4T^4)+O(T^{-5})$. 
We can also find an analytical expression for $\Delta \mathcal{F}(T)$ and $-(1/T)\partial_T\mathcal{D}_{A\to B}$ 
\begin{equation}
\begin{split}
\Delta\mathcal{F}(T)=&\left({J_x^2+J_y^2}\right)/(4T^4)+O(T^{-5})\\
-(1/T)\partial_T \mathcal{D}_{A\to B}(T)&=\left({J_x^2+J_y^2}\right)/(4T^4)+O(T^{-5}),
\end{split}
\label{D12}
\end{equation}
which agrees with Eq.~(\ref{eq:discordFisher}). 

We note that $\Delta \mathcal F$ does not depend on $B_1$, $B_2$, and $J_z$. {This can be intuitively understood since $J_x=J_y=0$ yields a classical Ising model}, where the Gibbs state is a classical state with zero quantum discord. 
In this case, Eq.~(\ref{eq:discordFisher}) is exact for any temperature as trivially $\Delta \mathcal{F}(T)=-(1/T)\partial_T \mathcal{D}_{A\to B}(T)=0$ at any temperature. 
{The other case for $\Delta \mathcal{F}(T)=-(1/T)\partial_T \mathcal{D}_{A\to B}(T)$ to be exact at any temperature is $B_1=B_2=0$ and either $J_y=0$ or $J_x=0$. In this case, we can obtain $\Delta \mathcal{F}(T)=-(1/T)\partial_T \mathcal{D}_{A\to B}(T)={J_k^2~{\rm sech}^2\left(\frac{J_k}{2T}\right)}/(4T^4)$, $k=x$ or $y$ for $J_y=0$ or $J_x=0$.}

\begin{figure}
\centering
\subfigure{
\includegraphics[width=0.21\textwidth]{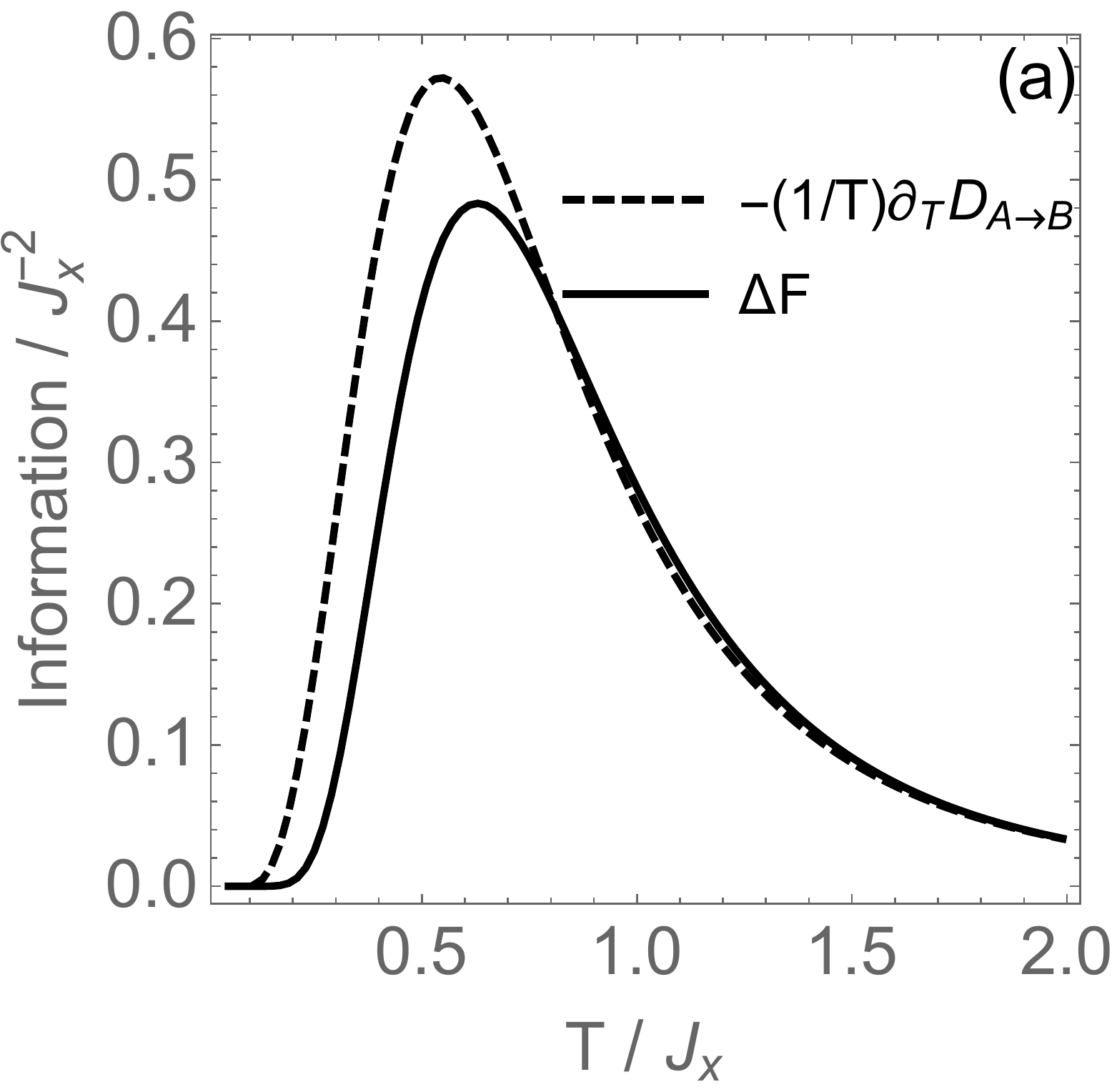}
}
\subfigure{
\includegraphics[width=0.2\textwidth]{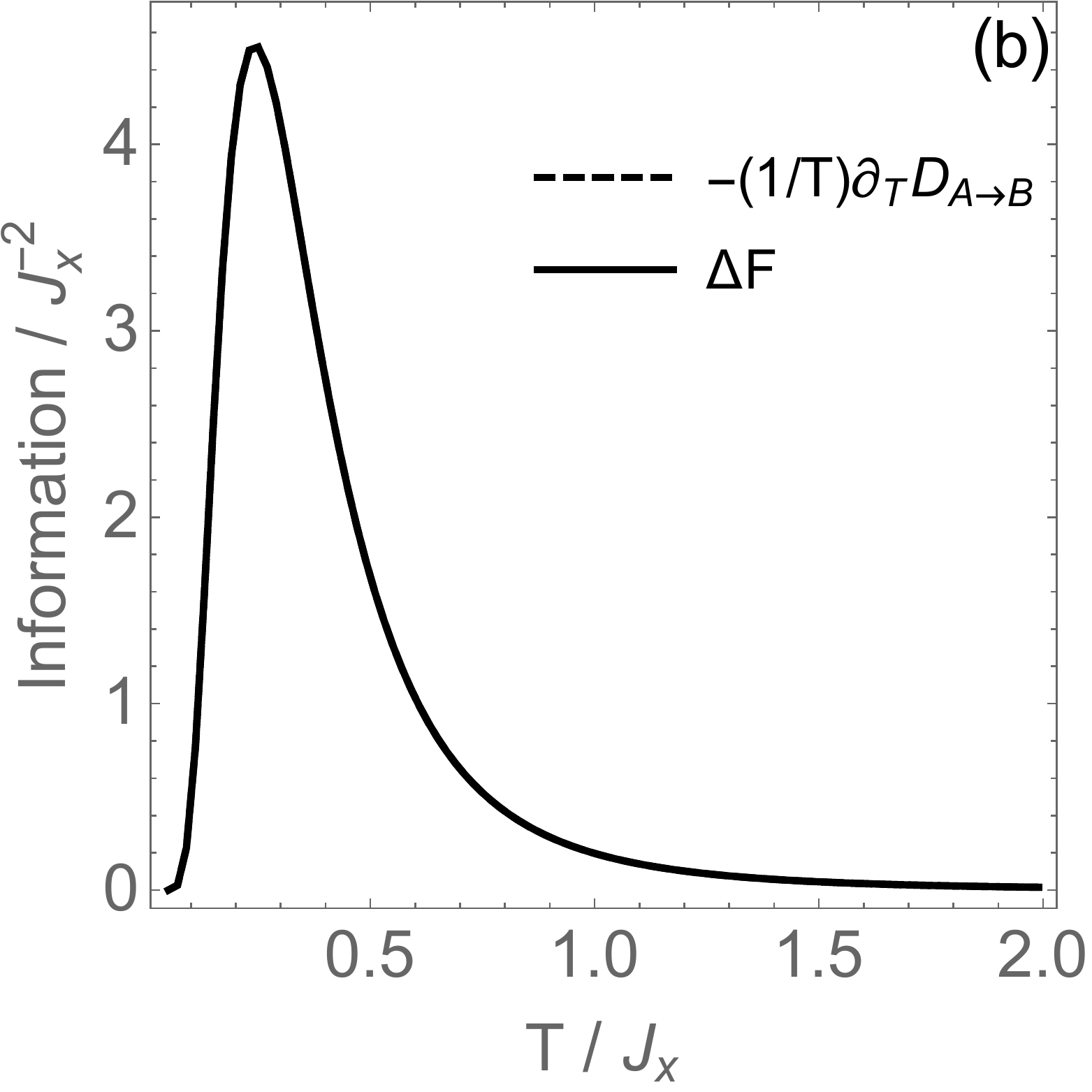}
}
\caption{$\Delta \mathcal{F}$ and $-(1/T)\partial_T \mathcal{D}_{A\to B}$, for a Heisenberg system with two qubits at (a) $B_1/J_x=3, B_2/J_x=1, J_z/J_x=2,J_y/J_x=1$ and (b) $B_1=B_2=0, J_z/J_x=2,J_y=0$.
}
\label{fig:D12}
\end{figure}

We can further numerically evaluate these quantities for arbitrary temperature, with results given in Fig.~\ref{fig:D12} for representative parameters. To understand the nontrivial parameter region better, since our model is symmetric between $J_x$ and $J_y$, without loss of generality, we fix $J_x$ and vary $J_y/J_x, B_1/J_x$, and  $B_2/J_x$. We find that for various parameters, at high temperature $\Delta\mathcal{F}(T)$ and $-(1/T)\partial_T \mathcal{D}_{A\to B}(T)$ agree well. At intermediate and low temperature, however, we find that the behavior of the quantities depends strongly on the system parameters. The relationship between $\Delta\mathcal{F}(T)$ and nonclassical correlation at low temperature is still an open problem.

\section{Multipartite Systems}
We now extend these ideas to multipartite systems. Suppose that we have a finite dimensional system composed of $N$ subsystems. We index each subsystem with an integer $1\le k \le N$. We want to quantify the difference in QFI between the sequential greedy measurement scheme on each subsystem and the global measurement.
We can sequentially apply the bipartite result in Eq.~(\ref{eq:discordFisher}) to derive the difference of QFI between the local and global schemes in the multipartite case. 

Let $\sigma_{1:N}\equiv \left(\sigma_1, \sigma_2,\cdots, \sigma_N\right)$, where $\sigma_k\in\{1,2,\cdots,N\}$, denote the measurement order of the local greedy scheme. 
At step $k=1$, there is no prior measurement results yet. By treating the system $\sigma_{1:N}$ as a bipartite composition of $\sigma_1$ and $\sigma_{2:N}$, Eq.~(\ref{eq:discordFisher}) gives the difference between global and LOCC QFI, i.e.
$
\mathcal{F}_{\sigma_{1:N}}
-\mathcal{F}_{\sigma_1\to \sigma_{2:N}}\simeq 
-(1/T)\partial_T \mathcal{D}_{\sigma_1\to \sigma_{2:N}}.
$
At step $2\le k\le N-1$, conditioned on previous measurement results $M_{1:k-1}\equiv \left(M_{\sigma_1},M_{\sigma_2},\cdots, M_{\sigma_{k-1}}\right)$, by treating the rest of system as a bipartite composition of $\sigma_k$ and $\sigma_{k+1:N}$, Eq.~(\ref{eq:discordFisher}) gives the difference between global and LOCC QFI, i.e.
$
\mathcal{F}_{\sigma_{k:N}|M_{1:k-1}}
-\mathcal{F}_{\sigma_k\to \sigma_{k+1:N}|M_{1:k-1}}\simeq 
-(1/T)\partial_T \mathcal{D}_{\sigma_k\to \sigma_{k+1:N}|M_{1:k-1}},
$
where 
$
\mathcal{F}_{\sigma_k\to \sigma_{k+1:N}|M_{1:k-1}}= \mathcal{F}_{\sigma_k|M_{1:k-1}}+\mathcal{F}_{\sigma_{k+1:N}|M_{1:k}}.
$

Now we consider the unconditional QFI $\mathcal{F}_{\sigma_k\to \sigma_{k+1:N}|\sigma_{1:k-1}}\equiv \sum_{M_{1:k-1}}P\left(M_{1:k-1}\right)\mathcal{F}_{\sigma_k\to \sigma_{k+1:N}|M_{1:k-1}}$, we have
$
\mathcal{F}_{\sigma_{k:N}|\sigma_{1:k-1}}-(\mathcal{F}_{\sigma_k|\sigma_{1:k-1}}+\mathcal{F}_{\sigma_{k+1:N}|\sigma_{1:k}})
\simeq 
-(1/T)\partial_T \mathcal{D}_{\sigma_k\to \sigma_{k+1:N}|\sigma_{1:k-1}}, 
$
where
$\mathcal{D}_{\sigma_k\to \sigma_{k+1:N}|\sigma_{1:k-1}}\equiv \sum_{M_{1:k-1}}P\left(M_{1:k-1}\right)\mathcal{D}_{\sigma_k\to \sigma_{k+1:N}|M_{1:k-1}}$.
By adding the equation above from $k=1$ to $k=N-1$ and noting that the difference in QFI is $\Delta \mathcal{F}_{\sigma_{1:N}}\equiv \mathcal{F}_{\sigma_{1:N}}-\sum_{k=1}^N \mathcal{F}_{\sigma_k|\sigma_{1:k-1}}$,
\be
\Delta \mathcal{F}_{\sigma_{1:N}}(T)
\simeq 
-(1/T)\partial_T \mathcal{D}_{\sigma_{1:N}}(T)+ O(T^{-5}),
\label{Theorem1}
\ee
where 
\be
\mathcal{D}_{\sigma_{1:N}}(T)=\sum_{k=1}^{N-1}\mathcal{D}_{\sigma_k\to \sigma_{k+1:N}|\sigma_{1:k-1}}(T),
\label{D_multi}
\ee
is a multipartite generalization of the bipartite diagonal discord defined in Eq.~(\ref{eq:discord}) with respect to the ordering $\sigma_{1:N}$. Therefore, Eq.~(\ref{Theorem1}) is valid for finite dimensional systems in the Gibbs state at high temperature.

The simplicity of this expression masks the fact that $\mathcal{D}_{\sigma_{1:N}}$ is complicated; since in each term $\mathcal{D}_{\sigma_k\to \sigma_{k+1:N}|M_{1:k-1}}$, the optimal measurement may depend on the previous measurement results $M_{1:k-1}$. 
We can still get further insight by considering systems where the optimal measurement is the same for all previous measurement results. 
Let $\Pi_j^{\sigma_k}\equiv |j\rangle_{\sigma_k}\langle j|$ denote the eigenbasis projection of $\rho_{\sigma_k}$ at step $k$; the optimal measurement must be $\pi_{\sigma_k}=\sum_j \Pi_j^{\sigma_k}$, yielding 
$
\mathcal{D}_{\sigma_k\to \sigma_{k+1:N}|\sigma_{1:k-1}}
= 
S\left(\pi_{\sigma_k}\circ \cdots \circ \pi_{\sigma_{1}}\left(\rho_{1:N}\right)\right)
-
S\left(\pi_{\sigma_{k-1}}\circ \cdots \circ \pi_{\sigma_{1}}\left(\rho_{1:N}\right)\right)
$, where $\circ$ denotes concatenation of operators and $\rho_{1:N}$ is the state of the entire system.

Note that all measurements $\pi_{\sigma_{N}}, \pi_{\sigma_{N-1}}, \cdots, \pi_{\sigma_{1}}$ commute with each other because they are on orthogonal support. Equation.~(\ref{D_multi}) simplifies to
\ba
\mathcal{D}_{\sigma_{1:N}}=\sum_{k=1}^N S\left(\rho_k\right)
-S\left(\rho_{1:N}\right),
\label{D_symm}
\ea
and the measurement order does not change the difference in QFI, because each of them commutes and does not depend on previous measurements.

\begin{figure}
\subfigure{
\includegraphics[width=0.21\textwidth]{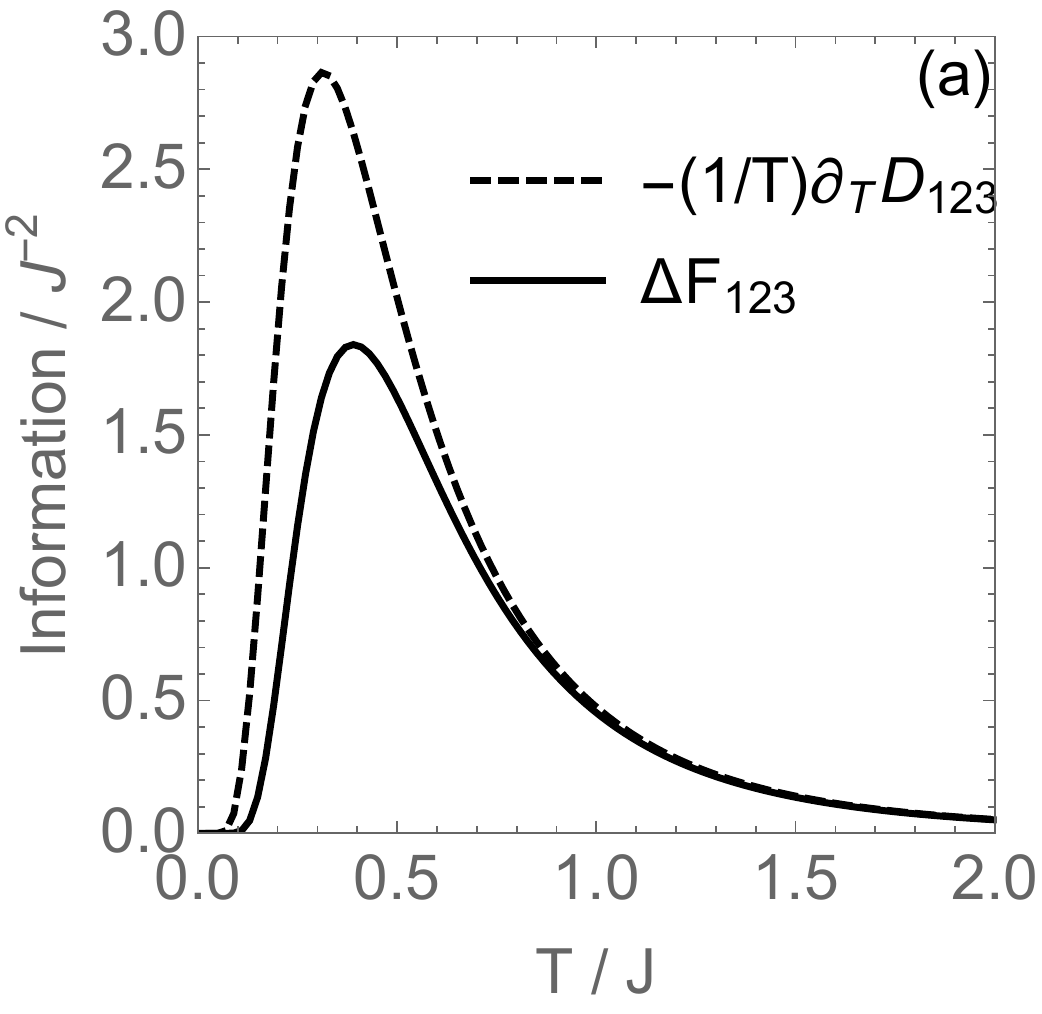}
}
\subfigure{
\includegraphics[width=0.21\textwidth]{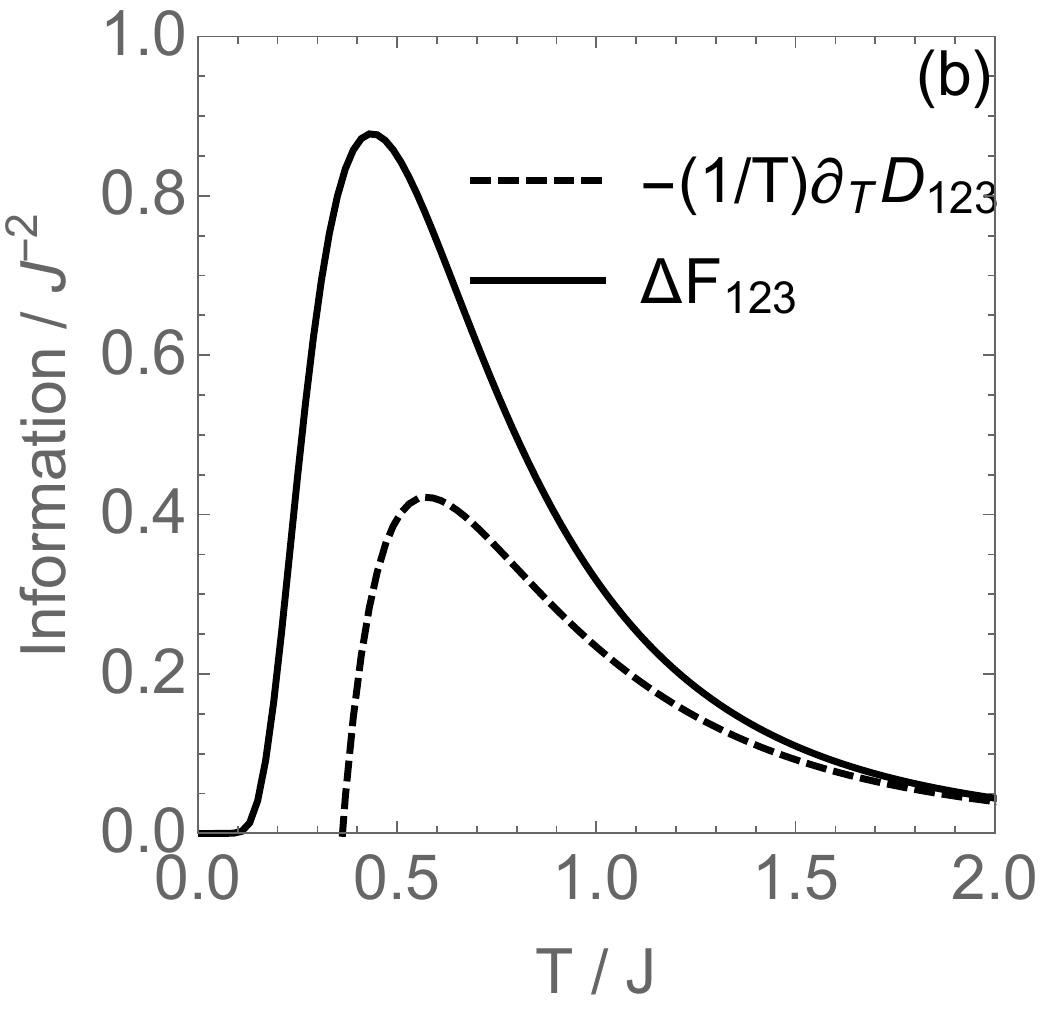}
}
\caption{$\Delta \mathcal{F}_{123}$ and $-(1/T)\partial_T \mathcal{D}_{123}$, for Heisenberg system with three qubits at (a) $B/J=1,\alpha=0.3$ and (b) $B/J=2,\alpha=0.3$. Note that the path denoted by subscript $132$ and $213$ have the same results.
}
\label{fig:D123}
\end{figure}

For example, consider the three-qubit Heisenberg system: $H=\frac{B}{2}\sum_{k=1}^3Z_k+\frac{J}{2}\sum_{k=1}^2\left(X_kX_{k+1}+Y_kY_{k+1}+\alpha Z_kZ_{k+1}\right)$. It has translational symmetry, and there are only three local measurement schemes to choose from: $1\to2\to3$, $1\to 3\to 2$, and $2\to3\to1$. However, we find that all three paths give the same $\Delta \mathcal{F}$ and diagonal discord. In the high-temperature limit we find $\Delta \mathcal{F}=-(1/T)\partial_T \mathcal{D}=J^2/T^4+O\left(T^{-5}\right)$ (see Fig.~\ref{fig:D123}).
Compared with Eq.~(\ref{D12}), we find that the loss is twice that of the two-qubit case, which is intuitive as there are two couplings.

More generally, if the Gibbs state is symmetric under permutation, the measurement order does not matter. However, even if $\mathcal{D}_{\sigma_{1:N}}$ is identical for all sequences $\sigma_{1:N}$, each measurement may still depend on previous measurement results. Still, if $N$ is large, we can show that feed forward is only required for the first few steps in a greedy local scheme. Indeed, according to the quantum de Finetti theorem~\cite{renner2007symmetry}, after a negligibly small number $K_1\ll N$ of measurements, the remaining $N-K_1$ subsystems becomes a mixture of independent and identically distributed states, i.e., $\rho_{1:N-K_1}\simeq \sum_{x} P_x \rho_x^{\otimes N-K_1}$. Because QFI is convex, we have $\mathcal{F}\left(T,\rho_{1:N-K_1}\right)\le \sum_{x} P_x \mathcal{F}\left(T,\rho_x^{\otimes N-K_1}\right)=\left(N-K_1\right) \sum_{x} P_x \mathcal{F}\left(T,\rho_x\right)$. This means that for the rest of the system, one can perform another $K_2\ll N$ number of measurements to determine $x$ and then perform the same local diagonal projection measurements on all $N-K_1-K_2$ parts in state $\rho_x$. 

\section{Conclusions and Outlook}
In conclusion, we have derived a relation between the diagonal discord and the LOCC QFI by comparing the global optimal measurement to a greedy local scheme in the high-temperature limit. We have proved that the diagonal discord quantifies the loss in temperature estimation precision due to performing a sequence of local measurements on subsystems of an arbitrary finite dimensional system. 
In other words, the nonclassical correlation other than entanglement, such as discord, can contribute to the precision enhancement in the temperature estimation.
This result demonstrates a close relation between nonclassical correlations and the ultimate precision limit in temperature estimation.

The relationship between precision loss in estimating temperature and diagonal discord could be potentially verified experimentally, exploiting nanoscale quantum devices. For example, recently, the local temperature of nanowires was measured~\cite{Idrobo18} through the electron energy gain and loss spectroscopy from room temperature to 1600 K. In general, predicting the precision loss in local measurements could guide experimentalists to select measurement protocols with the desired performance.

Although we focused on the high-temperature limit, the exploration of the finite- and low-temperature cases is an interesting open direction. {Indeed, for the two-qubit Heisenberg model, except for two analytical conditions for $-(1/T)\partial_T\mathcal{D}(T)=\Delta \mathcal{F}(T)~(\forall T)$ given in the main text, we also numerically observe that these two quantities are close to each other for various choices of the system parameters even at low temperature (see Appendix~\ref{app:morenumerical}).} We finally note that our derivation is only valid for finite-dimensional systems; the extension to infinite-dimensional systems is still open, due to the difficulty in the high-temperature expansion.

\acknowledgments
This work was supported in part by the U.S. Army Research Office through Grants No. W911NF-11-1-0400 and W911NF-15-1-0548 and by the NSF PHY0551153. 
QZ was supported by the Claude E. Shannon Research Assistantship.  We thank Anna Sanpera for helpful discussions. 

A.S. and Q.Z. contributed equally to this work.

\appendix

\section{Derivation of QFI of temperature estimation for the Gibbs state} 
\label{app:QFIGibbs}
Here, we review the derivation of QFI for the Gibbs state based on ~\cite{correa2015individual}.
Let $H$ be the Hamiltonian of a system thermalized at temperature $T$. Let $\beta$ be the inverse temperature, i.e., $\beta=1/T$, where we have set the Boltzmann constant as $k_B=1$. Then, the Gibbs state is given by
\begin{align*}
\rho(\beta)=\frac{1}{\mathcal{Z}_\beta}e^{-\beta H},
\end{align*}
where $\mathcal{Z}$ is the partition function:
\begin{align*}
\mathcal{Z}_\beta=\text{Tr}(e^{-\beta H}).
\end{align*}
Suppose that we have an error $\epsilon$ when estimating $\beta$. Then, the state with this error is given by:
\begin{align*}
\rho(\beta+\epsilon)=\frac{1}{\mathcal{Z}_{\beta+\epsilon}}e^{-(\beta+\epsilon)H}.
\end{align*}

QFI $\mathcal{F}(\beta)$ to estimate $\beta$ is defined by
\begin{align*}
\mathcal{F}(\beta)=-2\lim_{\epsilon\to 0}\frac{\partial^2}{\partial\epsilon^2}\mathbb{F}[\rho(\beta),~\rho(\beta+\epsilon)],
\end{align*}
where $\mathbb{F}[\rho(\beta),~\rho(\beta+\epsilon)]$ is the fidelity between $\rho(\beta)$ and $\rho(\beta+\epsilon)$
\begin{align*}
\mathbb{F}[\rho(\beta),\rho(\beta+\epsilon)]=\Big(\text{Tr}\sqrt{\rho^{1/2}(\beta)\rho(\beta+\epsilon)\rho^{1/2}(\beta)}\Big)^2.
\end{align*}

Now, let us calculate the fidelity first. The fidelity is given by
\begin{align*}
\begin{split}
\mathbb{F}[\rho(\beta),\rho(\beta+\epsilon)]&=\Big(\text{Tr}\sqrt{\rho^{1/2}(\beta)\rho(\beta+\epsilon)\rho^{1/2}(\beta)}\Big)^2\\
&=\frac{1}{\mathcal{Z}_{\beta}\mathcal{Z}_{\beta+\epsilon}}\Big(\text{Tr}\sqrt{e^{-\frac{1}{2}\beta H}e^{-(\beta+\epsilon)H}e^{-\frac{1}{2}\beta H}}\Big)^2\\
&=\frac{1}{\mathcal{Z}_{\beta}\mathcal{Z}_{\beta+\epsilon}}\Big(\text{Tr}\sqrt{e^{-(2\beta+\epsilon)H}}\Big)^2\\
&=\frac{1}{\mathcal{Z}_{\beta}\mathcal{Z}_{\beta+\epsilon}}\Big(\text{Tr}[e^{-(\beta+\frac{\epsilon}{2})H}]\Big)^2\\
&=\frac{\mathcal{Z}_{\beta+\frac{\epsilon}{2}}^2}{\mathcal{Z}_{\beta}\mathcal{Z}_{\beta+\epsilon}}.
\end{split}
\end{align*}

Before calculating the QFI, let us show the following fact:
\begin{align*}
\begin{split}
\lim_{\epsilon\to0}\frac{\partial}{\partial\epsilon}\mathcal{Z}_{\beta+\epsilon}=&-\lim_{\epsilon\to0}\text{Tr}[e^{-(\beta+\epsilon)H}H]=-\text{Tr}[e^{-\beta H}H]\\
\lim_{\epsilon\to0}\frac{\partial^2}{\partial\epsilon^2}\mathcal{Z}_{\beta+\epsilon}=&\lim_{\epsilon\to0}\text{Tr}[e^{-(\beta+\epsilon)H}H^2]=\text{Tr}[e^{-\beta H}H^2]\\
\lim_{\epsilon\to0}\frac{\partial}{\partial\epsilon}\mathcal{Z}_{\beta+\frac{\epsilon}{2}}=&-\lim_{\epsilon\to0}\text{Tr}\Big[e^{-(\beta+\frac{\epsilon}{2})H}\frac{H}{2}\Big]=-\frac{1}{2}\text{Tr}[e^{-\beta H}H]\\
\lim_{\epsilon\to0}\frac{\partial^2}{\partial\epsilon^2}\mathcal{Z}_{\beta+\frac{\epsilon}{2}}=&\lim_{\epsilon\to0}\text{Tr}\Big[e^{-(\beta+\frac{\epsilon}{2})H}\frac{H^2}{4}\Big]=\frac{1}{4}\text{Tr}[e^{-\beta H}H^2]\\
\end{split}
\end{align*}

For two functions $f(x)$ and $g(x)$, where $g(x)\neq0$, we have: 

\begin{align*}
\begin{split}
\frac{\partial^2}{\partial x^2}\frac{f^2}{g}=&2 \frac{\partial^2 f}{\partial x^2}\frac{f}{g}+\frac{2}{g}\Big(\frac{\partial f}{\partial x}\Big)^2\\
&-\frac{4f}{g^2}\Big(\frac{\partial f}{\partial x}\Big)\Big(\frac{\partial g}{\partial x}\Big)-\frac{\partial^2 g}{\partial x^2}\cdot\frac{f^2}{g^2}+\frac{2 f^2}{g^3}\Big(\frac{\partial g}{\partial x}\Big)^2.
\end{split}
\end{align*}

Therefore, if we define $x=\epsilon$, $f=\mathcal{Z}_{\beta+\frac{\epsilon}{2}}$, and $g=\mathcal{Z}_{\beta+\epsilon}$, we can obtain

\begin{align*}
\begin{split}
\lim_{\epsilon\to0}\frac{\partial^2}{\partial \epsilon^2}\mathbb{F}&=\lim_{\epsilon\to0}\frac{\partial^2}{\partial \epsilon^2}\frac{\mathcal{Z}_{\beta+\frac{\epsilon}{2}}^2}{\mathcal{Z}_{\beta}\mathcal{Z}_{\beta+\epsilon}}=\lim_{\epsilon\to0}\frac{1}{\mathcal{Z}_{\beta}}\frac{\partial^2}{\partial \epsilon^2}\frac{\mathcal{Z}_{\beta+\frac{\epsilon}{2}}^2}{\mathcal{Z}_{\beta+\epsilon}}\\
&=-\frac{1}{2}\text{Tr}\Big[\frac{e^{-\beta H}}{\mathcal{Z}_{\beta}}H^2\Big]+\frac{1}{2}\Big(\text{Tr}\Big[\frac{e^{-\beta H}}{\mathcal{Z}_{\beta}}H\Big]\Big)^2\\
&=-\frac{1}{2}\Big(\text{Tr}[\rho_{\beta}H^2]-(\text{Tr}[\rho_{\beta}H])^2\Big)\\
&=-\frac{1}{2}\Big(\langle H^2\rangle-\langle H\rangle^2\Big)=-\frac{1}{2}\delta H^2.
\end{split}
\end{align*}

Therefore, QFI becomes
\begin{align*}
\mathcal{F}(\beta)=-2\lim_{\epsilon\to0}\frac{\partial^2}{\partial \epsilon^2}\mathbb{F}=\delta H^2,
\end{align*}
which is the variance of the Hamiltonian. 
Therefore, with $M$ copies of the system, the variance of $\beta$ satisfies the following Cramer-Rao bound:
\begin{align*}
\epsilon^2\ge \frac{1}{M\mathcal{F}(\beta)}=\frac{1}{M\delta H^2}
\end{align*}
Since $\beta=1/T$, we have
\begin{align*}
\frac{\epsilon}{\delta T}=\frac{\delta\beta}{\delta T}=-\frac{1}{T^2},
\end{align*}
therefore, we can obtain
\begin{align*}
\delta T^2\ge\frac{T^4}{M\delta H^2}
\end{align*}
Therefore, we can find that QFI to estimate the temperature $T$ can be written as
\begin{align*}
\mathcal{F}(T)=\frac{\delta H^2}{T^4}=\frac{\mathcal{F}(\beta)}{T^4}.
\end{align*}
By definition, heat capacity $C(T)$ is given by
\begin{align*}
C(T)=\frac{1}{T^2}\delta H^2.
\end{align*}
QFI to estimate temperature $T$) for the Gibbs state becomes:
\begin{align*}
\mathcal{F}(T)=\frac{C(T)}{T^2}.
\end{align*}

Here, let us explain the reason why the energy measurement is the optimum for the Gibbs state. The measurement result is $\langle H\rangle(T)=\text{Tr}[\rho H]$, and the the variance is $\delta H^2=\langle H^2\rangle-\langle H\rangle^2$. In the single-shot scenario, estimation variance $\delta T$ can be written as
\begin{align*}
\delta T=\frac{\delta H}{|\partial_T \langle H\rangle|}.
\end{align*}
Here, note that for the Gibbs state, 
\begin{align*}
C(T)=\partial_T\langle H\rangle=\frac{(\delta H)^2}{T^2},
\end{align*}
we have $\delta T=T/\sqrt{C(T)}$, 
so that the variance of the temperature becomes
\begin{align*}
\delta T^2=\frac{T^2}{C(T)}.
\end{align*}
Since QFI is $\mathcal{F}(T)=\frac{C(T)}{T^2}$, we can find that
\begin{align*}
\delta T^2=\delta T_{\text{min}}^2=\frac{1}{\mathcal{F}(T)},
\end{align*}
which indicates that the energy measurement is the optimum.

\section{Derivation of $\mathcal{F}_{A\to B}(T)$}
\label{app:Markovian}
 QFI is simply the classical Fisher information over the optimal quantum measurement.
Consider an arbitrary consecutive measurement result $\left(X,Y\right)$ on $A$ and $B$. Despite the quantum nature of the measurement, a classical derivation suffices. The joint distribution is a Markovian chain $X\to Y$ and thus the joint distribution is
\be
P_{X,Y}\left(x,y;T\right)=P_X\left(x;T\right) P_{Y|X}\left(y|x;T\right)\nonumber.
\ee

We consider the most general scenario where the measurement result is continuous. The discrete case in the main text can be seen as a special case.
The greedy local measurement scheme has constrained Fisher information
\begin{widetext}
\begin{align}
\mathcal{F}_{A\to B}(T)
&=\int dx dy P_{X,Y}\left(x,y;T\right) \left(\partial_T \ln P_{X,Y}\left(x,y;T\right) \right)^2
\nonumber
\\
&=\int dx dy P_X\left(x;T\right) P_{Y|X}\left(y|x;T\right)\left(\partial_T \ln P_X\left(x;T\right)+\partial_T\ln P_{Y|X}\left(y|x;T\right) \right)^2
\nonumber
\\
&=\int dx dy P_X\left(x;T\right) P_{Y|X}\left(y|x;T\right)
\left[ \left(\partial_T \ln P_X\left(x;T\right)\right)^2+\left(\partial_T\ln P_{Y|X}\left(y|x;T\right) \right)^2\right]
\nonumber
\\
&=
\int dx P_X\left(x;T\right) 
 \left(\partial_T \ln P_X\left(x;T\right)\right)^2+
\int dx P_X\left(x;T\right) \int dy P_{Y|X}\left(y|x;T\right)\left(\partial_T\ln P_{Y|X}\left(y|x;T\right) \right)^2
\nonumber
\\
&=\mathcal{F}_A(T)+\mathcal{F}_{B|A}(T)\nonumber.
\end{align}
\end{widetext}
Note the cross term $\partial_T \ln P_X\left(x;T\right) \partial_T\ln P_{Y|X}\left(y|x;T\right)$ integrates to zero in the second step. To obtain the last line, we have used the fact that the greedy local measurement scheme saturates the local QFI on $A$.

\section{Proof of Eq.~(\ref{eq:prob})} 
\label{app:prob}
Here, we consider $p_{j*}(T)$. Let $d_{A}$ and $d_B$ be the dimensions of the subsystems $A$ and $B$, respectively, and the dimension of the total system $d_{AB}$ is written as $d_{AB}=d_Ad_B$. 
By definition, {from Eq.~(\ref{eq:highTexpansion}), in the high-temperature limit,} we can obtain
{
\begin{align*}
p_{j*}(T)&=\text{Tr}\Big[(\Pi_{j*}^A\otimes\openone_B)\rho_{AB}(\Pi_{j*}^A\otimes\openone_B)\Big]=\frac{1}{d_A}+O(T^{-1}),
\end{align*}
where we use the fact that $\text{Tr}[\openone_B]=d_B$.}


Therefore, we have
\begin{align*}
\partial_T p_{j*}(T)=O(T^{-2}). 
\end{align*}
Also, because $\sum_{j*}p_{j*}(T)=1$, we have 
\begin{align*}
\sum_{j*} \partial_T p_{j*}(T)=0
\end{align*}
and also the order of magnitude of the entropy is given by
\begin{align*}
S_{B|\Pi_{j}^A}(T)=\ln \left( d_B\right)+O(T^{-2}).
\end{align*}
Therefore, we can obtain
\begin{align*}
\begin{split}
\frac{1}{T}&\sum_{j*}\partial_T p_{j*}(T)S_{B|\Pi_{j*}^A}(T)\\
&=\frac{1}{T}\sum_{j*}\partial_T p_{j*}(T) \ln\left(d_B\right)+O(T^{-5})=O(T^{-5}).
\end{split}
\end{align*}

\section{Proof of Eq.~(\ref{eq:fisherorder})}
\label{app:fisherorder}
Let $H$ be the Hamiltonian for the finite-dimensional system. Then, the partition function can be written as
\begin{align*}
\mathcal{Z}=\text{Tr}[e^{-\beta H}]=\sum_{k=1}^{d}e^{-\beta h_k}.
\end{align*}
where $d$ is the dimension of the Hamiltonian (i.e., the number of eigenvalues of $H$), and $\{h_k\}_{k=1}^{d}$ are the eigenvalues of the Hamiltonian $H$. Then, the heat capacity $C(\beta)$ at high temperature ($\beta\ll1$) can be written as:
\begin{align*}
\begin{split}
C(\beta)&=\Big[~ \frac{1}{d}\sum_{k=1}^dh_k^2-\Big(\frac{1}{d}\sum_{k=1}^{d}h_k\Big)^2~\Big]\beta^2+O(\beta^3)\\
&=\delta h^2\beta^2+O(\beta^3),
\end{split}
\end{align*}
where 
\begin{align*}
\delta h^2=\frac{1}{d}\sum_{k=1}^d h_k^2-\Big(\frac{1}{d}\sum_{k=1}^{d}h_k\Big)^2
\end{align*}
is the variance of the eigenvalues. 
Since $\beta=1/T$, we have
\begin{align*}
C(T)=\frac{\delta h^2}{T^2}+O(T^{-3}).
\end{align*}
For the Gibbs state, the QFI of estimating temperature is
\begin{align*}
\mathcal{F}(T)=\frac{C(T)}{T^2}.
\end{align*}
Therefore, the order of magnitude of $\mathcal{F}(T)$ is 
\begin{align*}
\mathcal{F}(T)=O(T^{-4}).
\end{align*}

In our approach, in the high-temperature limit, the subsystem can be regarded as the Gibbs state; $\mathcal{F}_A(T)$, $\mathcal{F}_{B|A}(T)$, and $\mathcal{F}_{AB}(T)$ all have the order of magnitude $O(T^{-4})$. 
Therefore, if the greedy local method is not asymptotically optimal at high temperature, i.e., 
\begin{align*}
\lim_{T\to\infty}\frac{\Delta \mathcal{F}(T)}{\mathcal{F}(T)}>0,
\end{align*}
then we have 
\begin{align*}
\Delta \mathcal{F}(T)=O(T^{-4}),
\end{align*}
which shows that $\Delta \mathcal{F}(T)$ is more dominant in the high-temperature limit, i. e.
\begin{align*}
\Delta\mathcal{F}(T)\gg\frac{1}{T}\sum_{j*}\partial_T p_{j*}(T)S_{B|\Pi_{j*}^A}(T).
\end{align*}

\newpage
\section{More numerical results at low temperature}
\begin{figure}[http!]
\centering
\subfigure{
\includegraphics[width=0.45\textwidth]{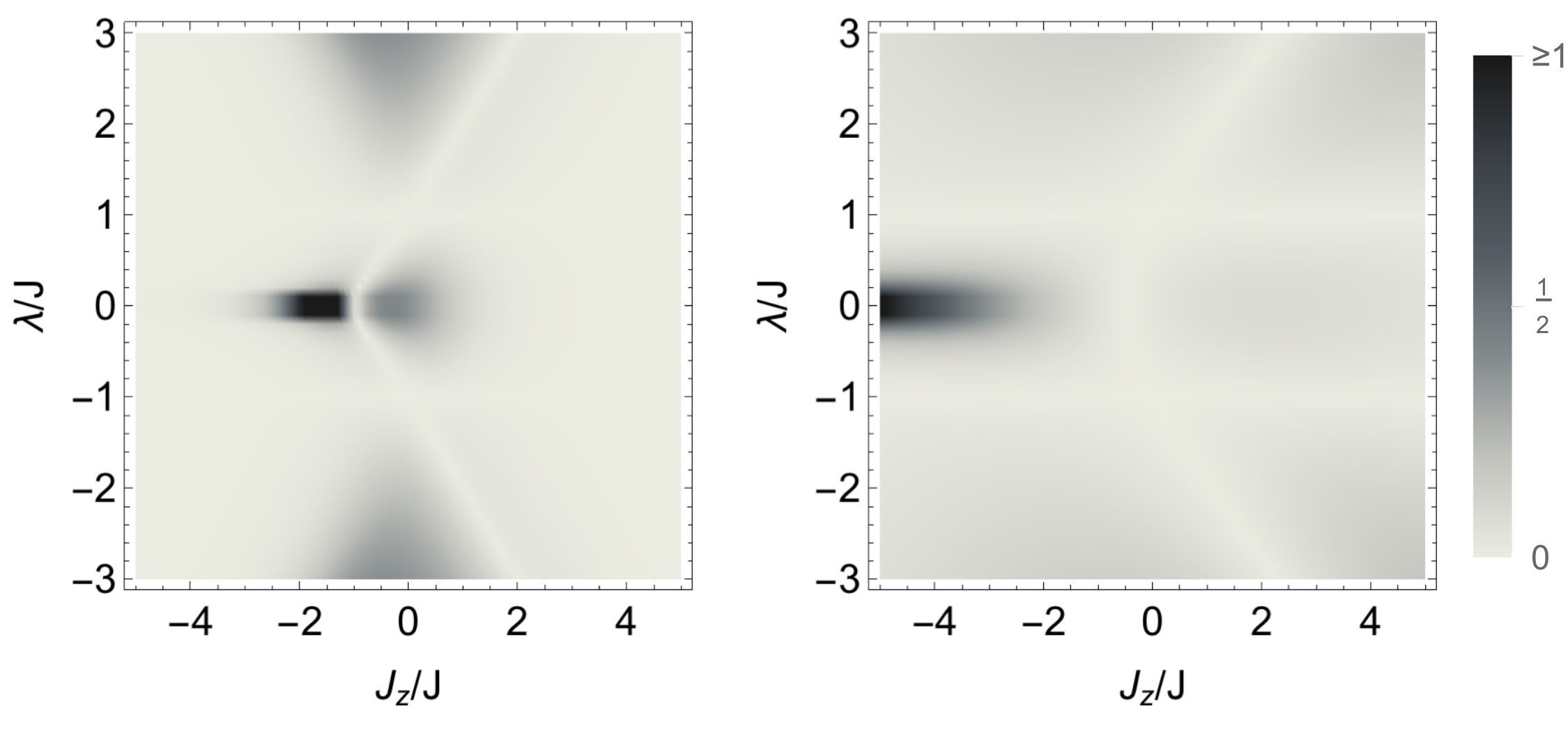}
}
\caption{ $|\left(\Delta \mathcal{F}+(1/T)\partial_T \mathcal{D}_{A\to B}\right)/\left(\Delta \mathcal{F}-(1/T)\partial_T \mathcal{D}_{A\to B}\right)|$. (a) $T/J=0.4$. Note that the increase of relative error at the edges is due to larger coupling amplitude making $T/|J\pm \lambda|$ smaller. (b) $T/J=2$. 
\label{fig:D12_relative}
}
\end{figure}

We consider the two-qubit Heisenberg interaction Hamiltonian in the absence of external fields
\begin{align*}
H=\left(1/2\right)\left(\left(J+\lambda\right)X_AX_B+\left(J-\lambda\right)Y_AY_B+{J_z}Z_AZ_B\right)
\end{align*}
To demonstrate the consistency between $\Delta\mathcal{F}(T)$ and $-(1/T)\partial_T \mathcal{D}_{A\to B}(T)$, we plot the relative difference $|\left(\Delta \mathcal{F}+(1/T)\partial_T \mathcal{D}_{A\to B}\right)/\left(\Delta \mathcal{F}-(1/T)\partial_T \mathcal{D}_{A\to B}\right)|$
 in Fig.~\ref{fig:D12_relative}. We see that except for a small region, the relative difference is small for both $T/J=0.4$ and $T/J=2$.

\label{app:morenumerical}

\bibliographystyle{apsrev4-1}
\bibliography{Biblio.bib}

\end{document}